\documentclass{article} %
\usepackage{nips11submit_e,times}
\usepackage{subfigure}
\usepackage{graphicx}
\usepackage{url}
\usepackage{common}

\title{Supporting the Curation of Twitter User Lists}

\author{Derek Greene, Fergal Reid \\ 
School of Computer Science \& Informatics,\\ 
University College Dublin, Ireland\\ 
\tt{\normalsize derek.greene@ucd.ie, fergal.reid@gmail.com} \\
\And
Gavin Sheridan \\
Storyful \\
Dublin, Ireland \\
\texttt{gavin.sheridan@storyful.com} \\
\And
P\'{a}draig Cunningham \\ 
School of Computer Science \& Informatics,\\ 
University College Dublin, Ireland\\ 
\tt{\normalsize padraig.cunningham@ucd.ie}
}

\nipsfinalcopy

\begin{document}
\maketitle

\begin{abstract}

Twitter introduced \emph{lists} in late 2009 as a means of curating tweets into meaningful themes. Lists were quickly adopted by media companies as a means of organising content around news stories. Thus the curation of these lists is important, they should contain the key information gatekeepers and present a balanced perspective on the story. Identifying members to add to a list on an emerging topic is a delicate process. From a network analysis perspective there are a number of \emph{views} on the Twitter network that can be explored, \eg followers, retweets mentions etc. We present a process for integrating these views in order to recommend authoritative commentators to include on a list. This process is evaluated on manually curated lists about unrest in Bahrain and the Iowa caucuses for the 2012 US election. 
\end{abstract}

\section{Introduction}
\label{sec:intro}
Media outlets that leverage the content produced by users of social media sites can now break or cover stories as they evolve on the ground, in real time (e.g. videos, photographs, tweets). However, a signiÞcant issue arises when trying to (a) identify content around a breaking news story in a timely manner (b) monitor the proliferation of content on a certain news event over a period of time, and (c) ensure that content is reliable and accurate. Storyful\footnote{\url{http://www.storyful.com}} is a social media news agency established in 2009 with the aim of filtering news, or newsworthy content, from the vast quantities of noisy data that streams through social networks. To this end, Storyful invests significant time into the manual curation of content on social media networks, such as Twitter and YouTube. In some cases this involves identifying ``gatekeepers'' who are prolific in their ability to locate, filter and monitor news from eyewitnesses.

Twitter users can organise the users they follow into Twitter \emph{lists}. Storyful maintains lists of users relevant to a given news story, as a means of monitoring breaking news related to that story. Often these stories generate community-decided hashtags (\eg \emph{\#occupywallstreet}) -- but even with small news events, using such hashtags to track the evolution of a story becomes difficult. Spambots quickly intervene, while users with no proximity (in space, time or expertise) to the news story itself drown out other voices.
Manual curation of lists is one way to overcome this problem, but is time consuming, and risks incomplete coverage.
In order to support the list curation process, we propose methods for identifying the important users that form the ``community'' around a news story on Twitter. Specifically, given a small seed list of users supplied by a domain expert, we are interested in using network analysis techniques to expand this set to produce a user list that provides comprehensive coverage of the story.
The motivation is that the members of this list will provide additional valuable content relating to the story.

A number of authors have considered the related problem of producing personal recommendations for additional users to follow on Twitter, either by following user links or performing textual analysis of tweet content. Hannon \etal \cite{hannon10twitter} proposed a set of techniques for producing personal recommendations on users to follow, based on the similarity of the aggregated tweets or ``profiles'' of users that are connected to the ego in the Twitter social graph. Such techniques have primarily relied on a single view of the network to produce suggestions. However, we can view the same Twitter network from a range of different perspectives.
 For instance, Conover \etal \cite{conover11polar} performed an analysis of Twitter data based on references to other Twitter screen names in a tweet, while researchers have also looked at the diffusion of content via \emph{retweets} to uncover the spread of memes and opinions on Twitter \cite{conover11polar,rat11truthy}. The idea is that both \emph{mentions} and \emph{retweets} provide us with some insight of the differing interactions between microblogging users.

In \refsec{sec:methods} we describe a set of recommendation criteria {\bf{and}} network exploration methods used to support user list curation on Twitter. Rather than using a single view of the network to produce recommendations, we employ a \emph{multi-view} approach that produces user rankings based on different graph representations of the Twitter network surrounding a given user list, and combines them using an SVD-based aggregation approach \cite{wu10recsys}. Information from multiple views is also used to control the exploration of the Twitter network -- this is an important consideration due to the limitations surrounding Twitter data access. To verify the accuracy of the resulting recommendations, in \refsec{sec:eval1} we describe experiments performed on a previously-curated Twitter list relating to coverage of the Iowa caucuses in advance of the 2012 US Presidential Election. In \refsec{sec:eval2} we investigate whether a ``silo'' effect arises in cases where a user list is expanded from an initial seed list with a strong bias towards a particular perspective on a story. We do this by evaluating the proposed recommendation techniques on subsets of a previously-curated list covering the current political situation in Bahrain. This study motivates further work in this area, which is discussed in \refsec{sec:conc}.

\section{Methods}
\label{sec:methods}

\subsection{Bootstrapping}

We now describe our proposed system for supporting user list curation. The initial input to the system is a seed list of one or more users that have been manually labelled as being relevant to a particular news story.  Once a seed list has been supplied, the first operation of the system involves a \emph{bootstrapping phase}, which retrieves follower ego networks around all seed list members. Other information regarding these users is also retrieved -- such as user list membership information and a limited number of tweets. The extent of the exploration process can be controlled by setting an upper limit for the number of links to follow and tweets to retrieve -- these parameters control the trade-off between network exploration depth and the number of queries required. The latter is an important consideration, not only in terms of running time, but also due to the fact that Twitter employs a quota system that limits the number of permitted API queries that can be made per hour.

After the bootstrap phase, the system will have two disjoint lists for the news story. The \emph{core set} contains curated Twitter accounts, initially this corresponds to the members of the seed list. The \emph{candidate set} contains Twitter accounts that are not in the core set, but exist in the wider network around the core -- some of these users may potentially be relevant for curation, while others will be spurious. Initially this will consist of the new non-seed users that were found during the bootstrap phase.

\subsection{Recommending Users}
\label{sec:rec}

In the subsequent \emph{recommendation phase}, a ranked list of the $r$ top users from the candidate set is produced. %
Firstly, we produce individual rankings using a number of criteria applied to different graphs, each representing a different view of the same network. The motivation is that each view potentially captures a different aspect of the relations between Twitter users around a given news story. We construct four different views:
\begin{enumerate}
\item \emph{Core friend graph}: This is a directed graph which contains nodes representing all users in the core set, along with the non-core users who they follow. 
\item \emph{Core mention graph:} As an alternative network view, we analyse the non-core users mentioned by the users in the core set. Specifically, an edge links from a core node A to a non-core node B if A has mentioned B in at least one tweet -- the weight of the edge corresponds to the number of tweets. The idea here is that this directed, weighted mention graph is a proxy for the dialogue between these Twitter users.
\item \emph{Core retweet graph:} We also analyse retweeting activity by core users involving tweets originally posted by  non-core users. This involves the construction of a weighted, directed  graph, where an edge links from core node A to non-core node B if A has retweeted B's tweets at least once -- the weight of the edge corresponds to the number of retweets. 
\item \emph{Weighted co-listed graph:} Another alternative view, which has not been widely explored in the literature, is to look at relations based on the aggregation of co-assignments to Twitter user lists. At an aggregate level, this could be regarded as a form of crowd-sourced curation, where the assumption is that related pairs of users will be more frequently assigned to the same list than users who have dissimilar to one another. Based on this idea, we construct a weighted, undirected graph as follows. For each user list that has been identified, we measure the overlap $w$ between the list's members and the core set using the Jaccard set similarity measure \cite{jaccard12index}. If $w > 0$ then, for each unique core/non-core pair of users in the user list, we create an edge between these two users with weight $w$. If an edge between the users already exists, we increment the weight on the edge by $w$.
\end{enumerate}
The criteria that we use on these graphs are as follows:
\begin{enumerate}
\item \emph{In-degree}: A simple approach for directed graphs is to look at the in-degree centrality of each Twitter user. For weighted graphs, we calculate the sum of the weights on incoming edges.
\item \emph{Normalised in-degree:} Using standard in-degree centrality can potentially lead to the selection of high-degree Twitter users who are not specialised in a particular geographic or topical area. Our solution has been to introduce a normalisation factor to reduce the impact of high degree nodes. The normalisation approach is similar to standard log-based TF-IDF term weighting functions that are widely applied in text mining to reduce the influence of frequently-occurring terms \cite{salton87tfidf}. The normalised follower count value for the user $u$ is defined as:
\begin{equation}
\textrm{nfc}(u) = 
\log{(seed\_followers(u))} 
\; \cdot \; \log \left( 
\frac{max\_followers}{all\_followers(u)}
\right)
\label{eqn:norm}
\end{equation}
where
\begin{itemize}
\item $seed\_followers(u)$ = the number of users in the core set that follow the user $u$.
\item $all\_followers(u)$ = the total number of all users following the user $u$ on Twitter.
\item  $max\_followers$ = a scaling factor, defined to be the largest number of Twitter followers among any of the core and non-core users.
\end{itemize}
\item \emph{HITS with priors}: The HITS algorithm, originally proposed by \cite{kleinberg99hits}, has been widely used to assign hub and authority scores to each mode in graph, depending upon its the topology. We can use the authority scores applied to a Twitter network to identify key users in that network. Since we wish to focus on authority relative to our pre-curated core list, we use the variation of HITS proposed by \cite{white03algorithms}, which introduces prior probabilities for each node. Specifically, each of the $m$ users in the core list is given an initial probability $1/m$, while the other non-core nodes are given an initial probability of 0.
\end{enumerate}
Naturally, certain criteria are only meaningful when applied to certain graphs. For the purpose of the evaluations described in this paper, we use the following five combinations:
\begin{itemize}
\item Normalised degree applied to the core friend graph.
\item HITS with priors applied to the core friend graph.
\item Weighted in-degree applied to the co-listed, mention, and retweet graphs.
\end{itemize}

\subsubsection{Combining Rankings}
The various graph/criterion combinations can potentially produce rankings of users that differ significantly. To combine rankings, we use SVD-based aggregation, which has previously been shown to be effective for this task \cite{wu10recsys}. We construct a matrix $\m{X}$ from the ranks (rather than the raw scores), with users on the rows and rankings on the columns. We then apply SVD to this matrix and extract the first left singular vector. The values in this vector provide aggregated scores for the users. By arranging these values in descending order, we can produce a final ranking of users. We select the top $r$ users to form our list of user recommendations.  Finally, we can also apply additional filtering of recommendations based on a minimum tweet count filter and a filter to remove users who have not tweeted within a given time period. 

After a set of recommendations has been generated, the ranked list of suggested users would be presented to a human curator, who could then select a subset to migrate to the core set (\ie to augment the existing Twitter user list). The use of a ``human in the loop'' in the proposed system resembles the role of the oracle in active learning algorithms for classification \cite{settles09review}.

\subsection{Network Exploration}
\label{sec:explore}

\begin{figure}[!t]
\centering
\includegraphics[width=\linewidth]{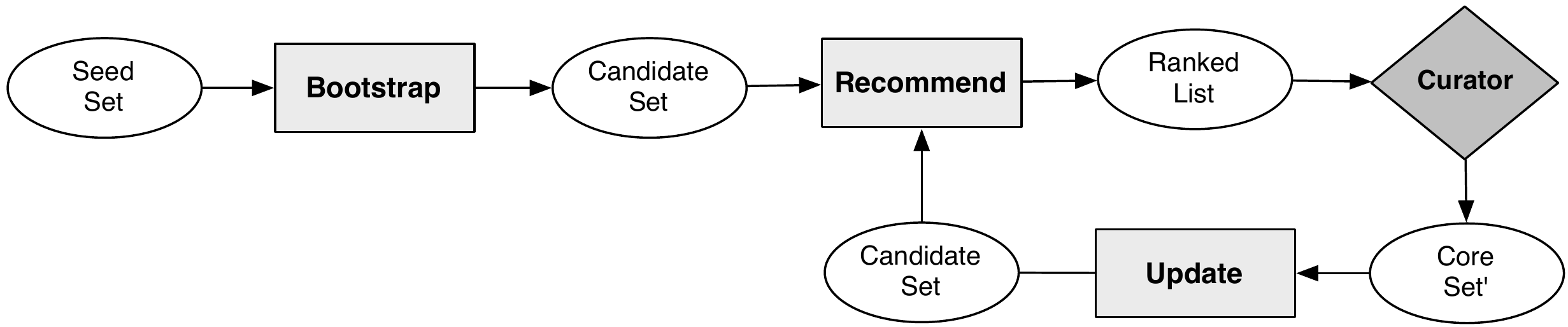}
\caption{Overview of curation support system, illustrating the workflow between the \emph{bootstrap}, \emph{recommendation}, and \emph{update} phases.}
\label{fig:workflow} 
\end{figure}

Once the core set has been modified, the system enters the \emph{update phase}, which modifies the current copy of the network to reflect (a) changes in membership of the core set, and (b) any changes in the Twitter network since the last update (\eg addition/removal of follower links, new tweets). Specifically, the network is explored using a process based both on the follower graph and also on tweet content:
\begin{itemize}
\item For the current core set, retrieve their friend/follower links, user list memberships, and recent tweets for all set members (\ie same process as in the bootstrap phase).
\item For the last set of recommended users who were not migrated to the core set, retrieve their friend/follower links, user list memberships, and recent tweets.
\item For the set of $m$ users who were most frequently mentioned in tweets posted by the core set, retrieve their friend/follower links, user list memberships, and recent tweets.
\end{itemize}
Again the extent of the exploration for each of the above can be controlled by setting maximum values for the number of links to follow and tweets to retrieve. Once the local copy of the network has been update, the data then feeds back to the recommendation phase and another iteration of the recommendation-selection-update process is executed. A visual overview of the complete curation system process is shown in \reffig{fig:workflow}.

\section{Case Study 1: Iowa}
\label{sec:eval1}

\subsection{Experimental Setup}
\begin{figure}[!t]
\centering
\subfigure[Initial core set]{
	\includegraphics[width=2.6in]{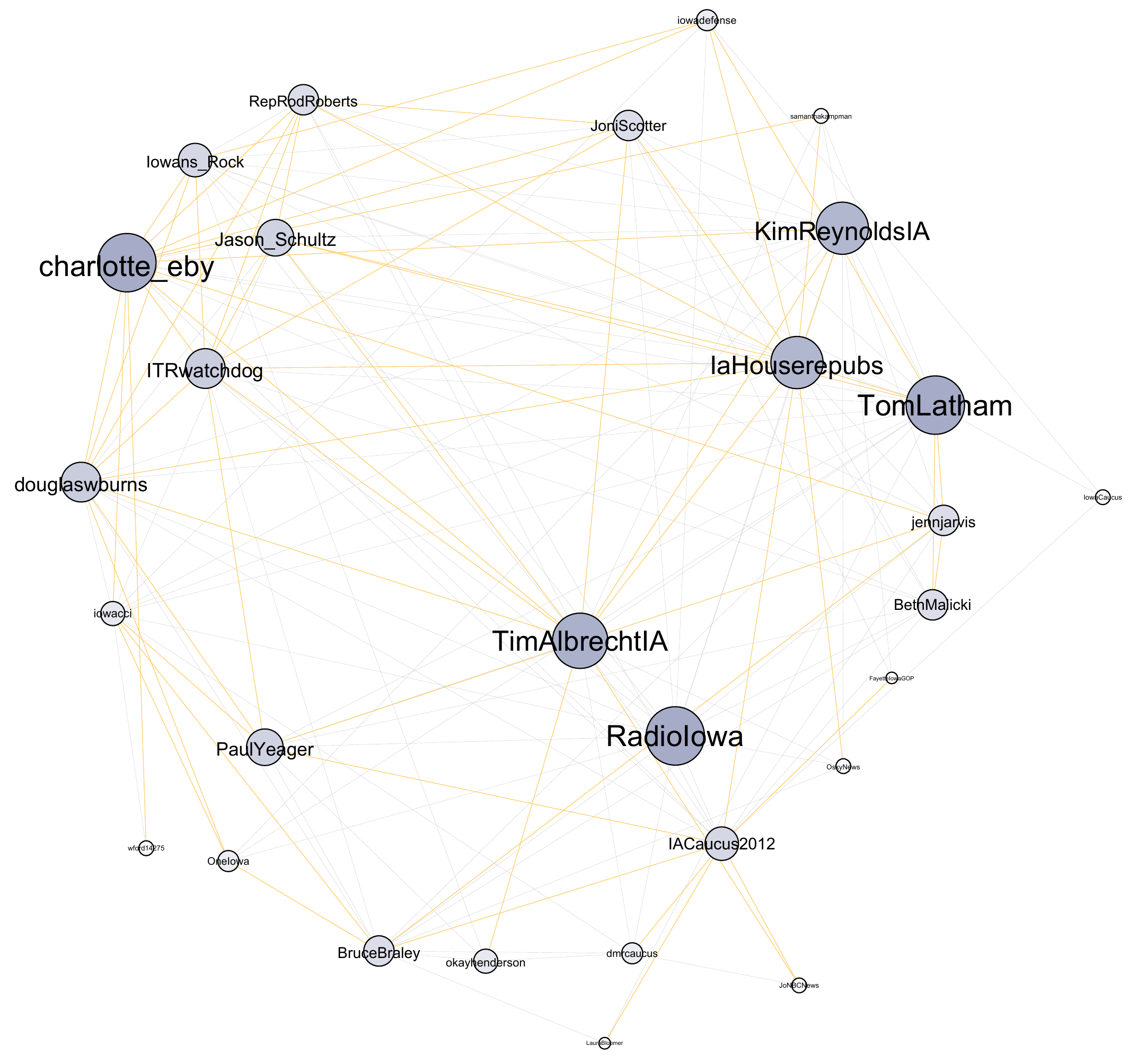}
	\label{fig:core1boot}
 }
 \hskip 0.3em
 \subfigure[Final core set]{
   \includegraphics[width=2.6in] {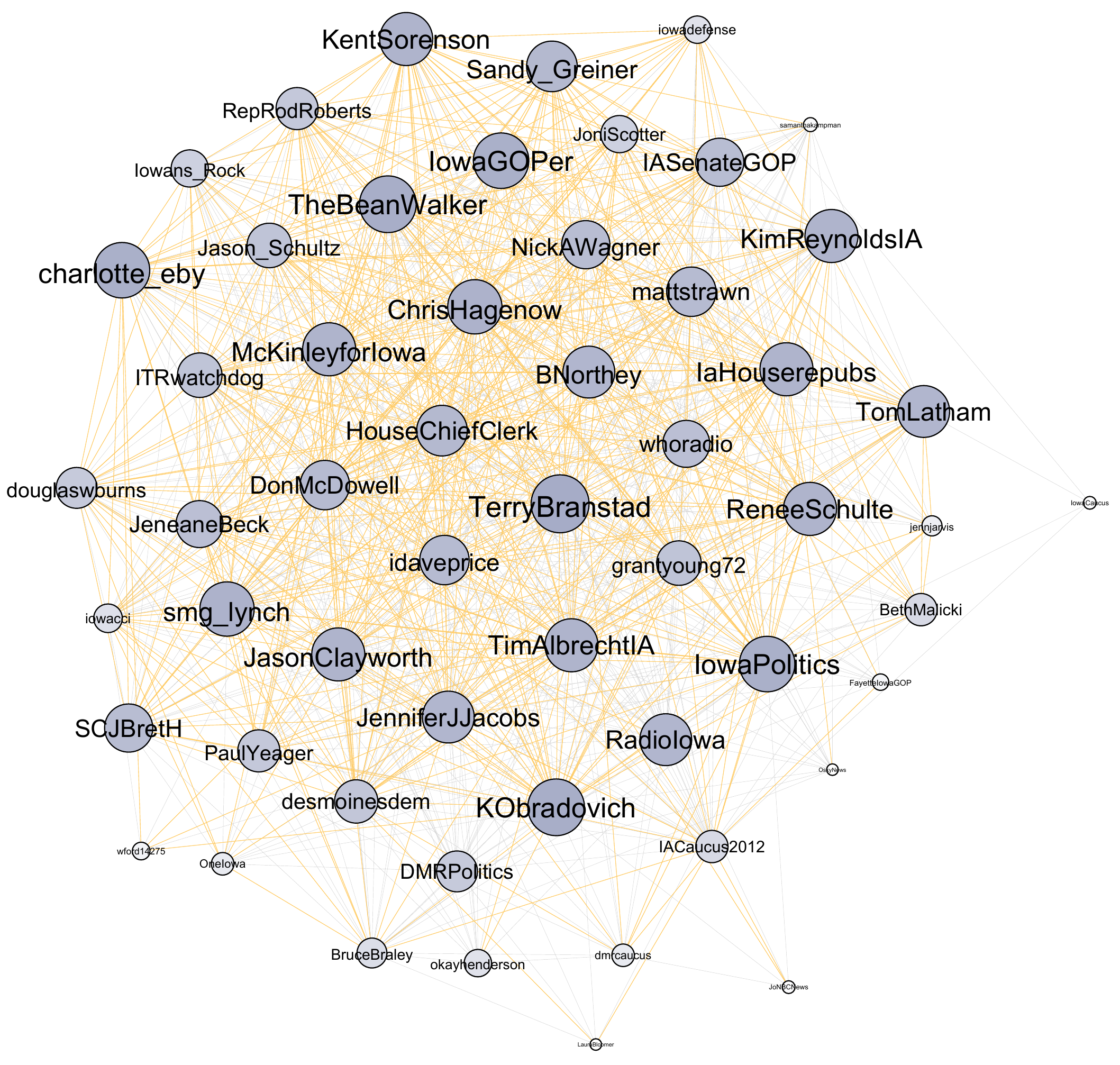}
   \label{fig:core1final}
 }
\label{fig:core1}
\caption{Induced subgraph of the follower graph for the core set members in the Iowa dataset \#1 after (a) the initial bootstrap phase, (b) six complete iterations. Larger nodes with a more saturated colour are indicative of nodes with a higher in-degree (\ie users with more followers within the core set). Highlighted edges indicate reciprocated follower links between users. Layout positions are preserved for both figures.}
\end{figure}

First, we evaluated the proposed recommendation system on a Twitter user list previously curated by Storyful, covering Iowa politics during the 2012 US Presidential Primaries\footnote{\url{http://twitter.com/#!/trailmix12/iowa}}. At the time of initial data collection -- 16 September 2011 -- this list contained 128 unique users. To evaluate the robustness of the user recommendation process, we use cross validation, randomly dividing the complete Iowa user list into four disjoint datasets, each containing 32 users. As an example, the subgraph induced by the core set on the follower graph of Iowa dataset \#1 is shown in \reffig{fig:core1boot} -- the positions of nodes were calculated using the force directed layout implementation provided by Gephi \cite{bastian09gephi}. 

In our experiments we applied the workflow shown in \reffig{fig:workflow} to each of the sets individually for six \emph{recommendation-selection-update} iterations after the initial bootstrapping phase. Note that no information was shared between the runs. The extent of network exploration during the update phase was controlled using the following constraints:
\begin{itemize}
\item A maximum of up to 1,000 friend/follower links were retrieved per user at a given iteration. 
\item A maximum of up to 1,000 user lists were retrieved per user at a given iteration.
\item A maximum of up to 1,000 tweets was retrieved per user at a given iteration.
\item Very high-degree users $> 50,000$ friends and/or 50,000 followers were filtered.
\end{itemize}
To generate recommendations, we used the views and criteria as described in \refsec{sec:methods}. We filtered the recommendations to remove those users who had not tweeted in the previous two weeks and/or those who had posted fewer than 25 tweets in total. At each iteration we generated $r=50$ recommendations -- by the final iteration, users were selected from a complete candidate set with average size of $\approx 62k$ users. At this stage, we had also collected an average $\approx 63k$ tweets and $\approx 138k$ follower links for each dataset. In place of a manual curator, after each complete iteration we automatically selected the top five highest ranked users (based on SVD aggregation) to add to the core set. The six iterations thus yielded 30 additional core users for each of the four sets. As an example, the final expanded core set for Iowa dataset \#1 is shown in \reffig{fig:core1final}. It is interesting to observe that several high-degree nodes were added to the core set, such as the user \emph{@TerryBranstad}, the official account of the Governor of Iowa.

\subsection{Discussion}
Next, to quantitatively validate the relevance of the recommendations produced by our proposed techniques, we use two measures that are frequently used in information retrieval tasks: \emph{precision} and \emph{recall}. The results for the four datasets across all six iterations are listed in \reftab{tab:iowa}. We observe that, in terms of recall, increasing the user list size by 30 accounts does not lead to a significant fall in precision -- average precision relative to the complete original list remains at 0.88 by iteration six. Meanwhile, recall increases steadily in all cases -- the average is 0.43. Note the maximum achievable recall by iteration six is 0.48 (\ie 62 out of the 128 users are returned), and is lower in previous iterations.

\begin{table}[!t]
\centering
\begin{tabular}{|c|ccccc|ccccc|}\hline
\bf Iteration & \multicolumn{5}{|c|}{\bf Precision} & \multicolumn{5}{|c|}{\bf Recall} \\ \hline
\bf  & \it Set 1 & \it Set 2 & \it Set 3 & \it Set 4 & \it Mean & \it Set 1 & \it Set 2 & \it Set 3 & \it Set 4 & \it Mean \\ \hline
\it 1 & 0.95 & 0.97 & 0.97 & 1.00 & \bf 0.97 & 0.27 & 0.28 & 0.28 & 0.29 & \bf 0.28 \\ 
\it 2 & 0.93 & 0.98 & 0.95 & 1.00 & \bf 0.96 & 0.30 & 0.32 & 0.31 & 0.33 & \bf 0.32 \\ 
\it 3 & 0.94 & 0.96 & 0.96 & 0.98 & \bf 0.96 & 0.34 & 0.35 & 0.35 & 0.36 & \bf 0.35 \\ 
\it 4 & 0.90 & 0.94 & 0.96 & 0.96 & \bf 0.94 & 0.37 & 0.38 & 0.39 & 0.39 & \bf 0.38 \\ 
\it 5 & 0.88 & 0.93 & 0.93 & 0.95 & \bf 0.92 & 0.39 & 0.41 & 0.41 & 0.42 & \bf 0.41 \\ 
\it 6 & 0.82 & 0.92 & 0.89 & 0.89 & \bf 0.88 & 0.40 & 0.45 & 0.43 & 0.43 & \bf 0.43 \\ \hline
\end{tabular}
\caption{Precision and recall scores for four randomly-selected subsets of the Storyful Iowa user list, for six complete  \emph{recommendation-selection-update} iterations. }
\label{tab:iowa}
\end{table}

In general, we observe that the Iowa user list studied here consists of a relatively homogeneous group of users pertaining to a story with a relatively narrow focus -- the users are predominantly Republicans involved in the Iowa caucuses. Therefore, unlike the study in \cite{conover11polar} which analysed Twitter relations across the entire country during 2010 US midterm elections, here a pronounced partisan divide is not evident. 

\section{Case Study 2: Bahrain}
\label{sec:eval2}

\subsection{Experimental Setup}
For our second study, we analyse a dataset with significantly different characteristics. As a seed list we use a Twitter list covering the current political situation in Bahrain which was also manually curated by Storyful\footnote{\url{http://twitter.com/#!/storyfulpro/bahrain}}. As of 27 September 2011, this list contained 51 users. A small number of these have a ``loyalist'' or ``pro-government'' stance, while the remaining users could be regarded as being either  ``non-loyalist'', or ``neutral'' observers with an interest in Bahrain. This natural division in the seed list raises an interesting question -- does starting with a seed list that takes a particular stance on a given news story lead to the construction of localised network ``silos'', which may lead an automated system to give biased user recommendations? 

To investigate this, we generate recommendations based on a seed list \emph{Bahrain-L} containing a subset of 14 users that have been putatively labelled as ``loyalist''. We ran four complete iterations using the same exploration constraints, filters, and selection mechanism used in the previous evaluation. This resulted in a core set containing 34 users, a candidate set of 51,114 users, 138,777 follower links, and 53,450 unique tweets.

\subsection{Discussion}

\begin{figure}[!t]
\centering
\subfigure[Initial core set]{
	\includegraphics[width=0.26\linewidth]{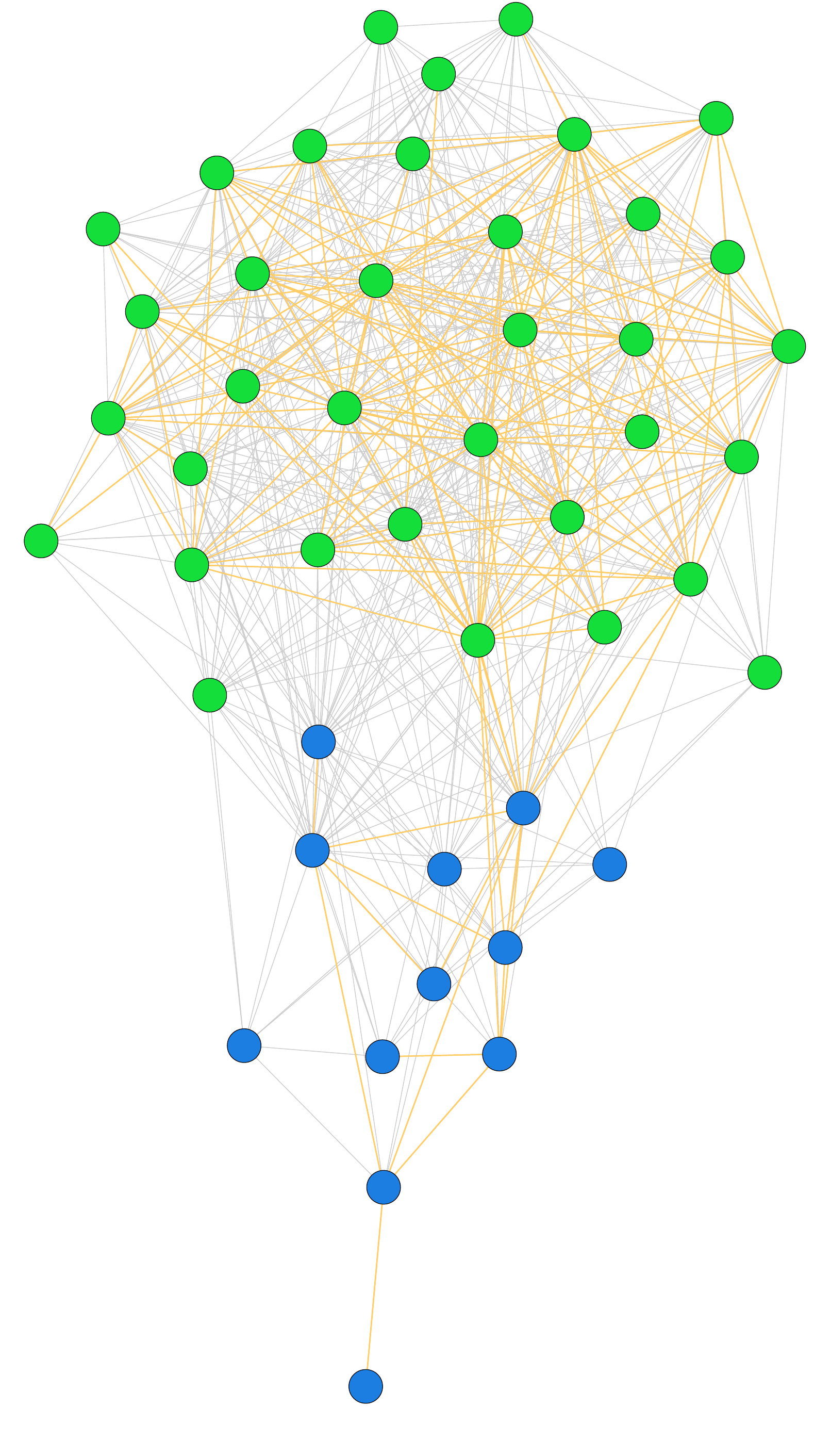}
	\label{fig:bah1}
 }
 \hskip 4em
 \subfigure[Final core set]{
   \includegraphics[width=0.348\linewidth] {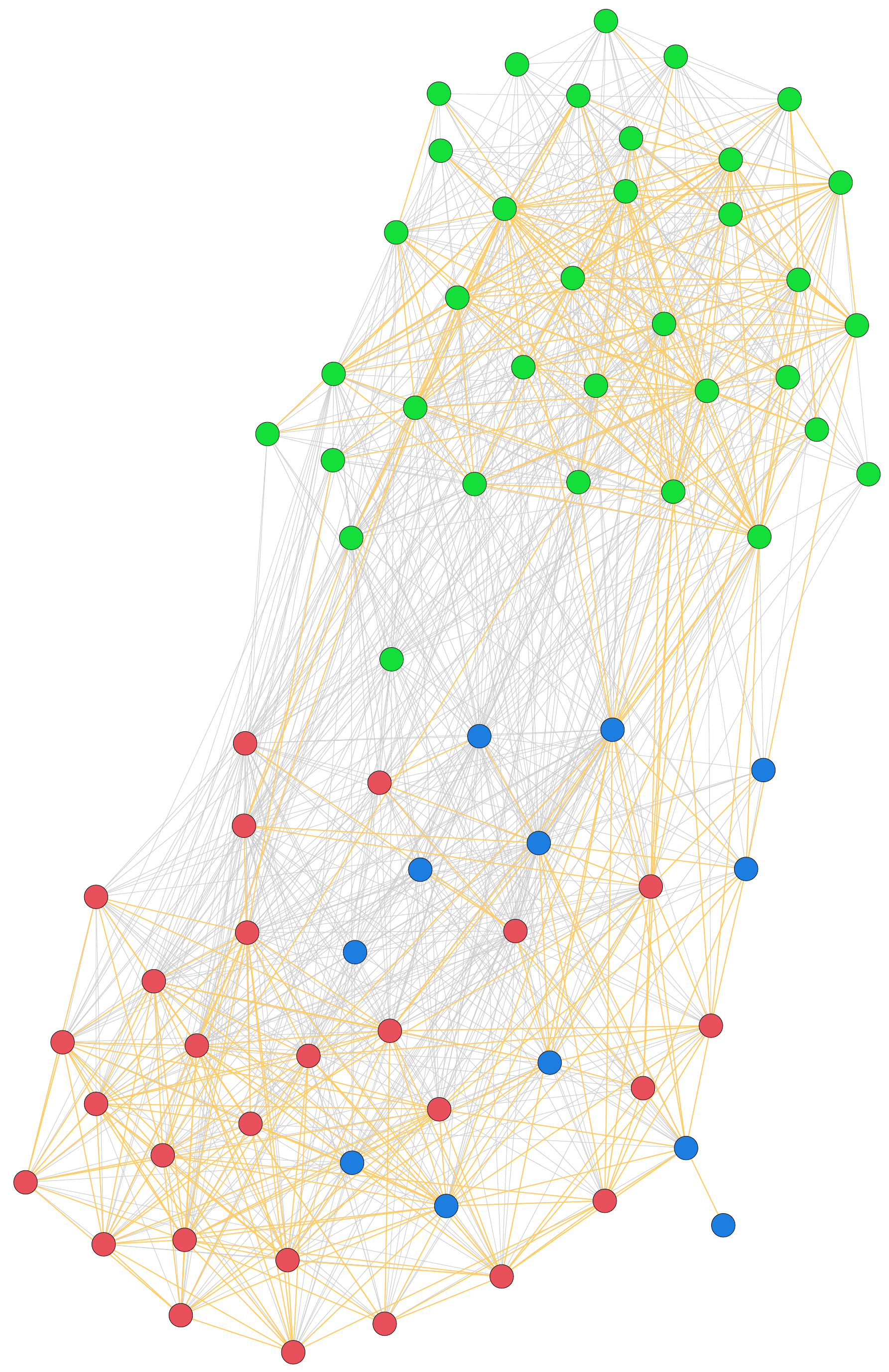}
   \label{fig:bah2}
 }
\label{fig:bah}
\caption{Follower graph for core set members in the Bahrain dataset after (a) the initial bootstrap phase, (b) four complete iterations. Blue nodes denote users in the original user list that are putatively labelled as ``loyalist'', while the remaining members of the user list are coloured green. The additional nodes that have been selected, based on recommendations using \emph{Bahrain-L} as a seed list, are coloured red.}
\end{figure}

\reffig{fig:bah1} shows the subgraph induced by the original complete curated list of 51 users on the follower graph -- the split between the ``loyalist'' users and the other users is evident from the positions calculated by force directed layout. In particular, the latter group of users form a densely connected core, while most of the ``loyalist'' nodes are not well-connected with the rest of the subgraph. \reffig{fig:bah2} shows a subgraph induced by the union of the curated list, with the set of nodes selected based on the recommendation process using \emph{Bahrain-L} alone as the seed set. We observe that none of the 37 ``non-loyalist''  nodes from the curated list were selected during the four iterations. In contrast, we see that the new users are closely connected with the other ``loyalist'' users, forming a second dense core. While we might expect this if recommendations were only generated based on follower links, recall that rankings based on mentions and retweets are also being aggregated to select new users. 
In fact, the addition of these rankings appears to further compound the ``silo'' effect which is evident from \reffig{fig:bah}. 

Our analysis suggests that there is little interaction on Twitter between users with differing stances on the political situation in Bahrain. On the one hand, this highlights weakness of the proposed recommendation techniques in the case of stories that are highly-polarised. Alternative criteria, which emphasise diversity over homogeneity, may provide a solution -- this is analogous to the attempts in active learning to identify diverse examples in order to widely cover the sample space \cite{settles09review}. On the other hand, these results also highlight the continued importance of the role of the curator in (a) selecting a suitably diverse seed list as a starting point, (b) actioning recommendations produced by the system.

\section{Conclusions}
\label{sec:conc}

In this paper we have proposed a comprehensive approach for automating aspects of the Twitter list curation process, based on novel network exploration and multi-view recommendation techniques. In the evaluation in \refsec{sec:eval1}, we showed that, using different starting subsets of a manually-curated list, we can recall the original human annotations while maintaining high precision.

Based on the observations made in \refsec{sec:eval2}, we suggest that the next major phase of our work will involve exploring the diffusion patterns of newsworthy multimedia resources (\eg links to images and videos) in the network surrounding a user list. For instance, identifying users who are frequently early in \emph{retweet chains} for such resources may help diversify user list recommendations in situations where the ``silo'' effect is pronounced, such as in the Bahrain case study. In future we also plan to apply the proposed recommendation and network exploration techniques beyond Twitter, looking at multiple views across several different social networks. A key issue here will be the generation of a reliable mapping between users on different networks.

\subsubsection*{Acknowledgments}
This work is supported by Science Foundation Ireland Grant No. 08/SRC/I140 (Clique: Graph \& Network Analysis Cluster). The authors thank Storyful for their participation in the evaluations performed in this paper.

\bibliographystyle{plain}
\bibliography{nips}

\end{document}